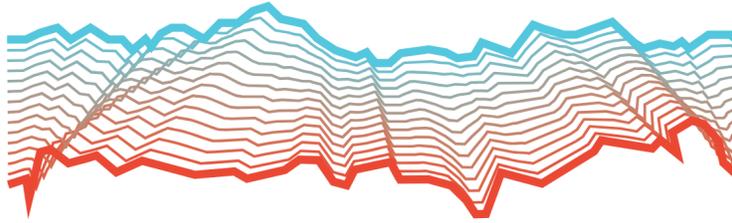

# Seismic Response of Yielding Structures Coupled to Rocking Walls with Supplemental Damping

M. Aghagholizadeh[1] and N. Makris[2]

**ABSTRACT**

Given that the coupling of a framing structure to a strong, rocking wall enforces a first-mode response, this paper investigates the dynamic response of a yielding single-degree-of-freedom oscillator coupled to a rocking wall with supplemental damping (hysteretic or linear viscous) along its sides. The full nonlinear equations of motion are derived, and the study presents an earthquake response analysis in term of inelastic spectra. The study shows that for structures with preyielding period $T_1 < 1.0$ s the effect of supplemental damping along the sides of the rocking wall is marginal even when large values of damping are used. The study uncovers that occasionally the damped response matches or exceeds the undamped response; however, when this happens, the exceedance is marginal. The paper concludes that for yielding structures with strength less than 10% of their weight the use of supplemental damping along the sides of a rocking wall coupled to a yielding structure is not recommended. The paper shows that supplemental damping along the sides of the rocking wall may have some limited beneficial effects for structures with longer preyielding periods (say $T_1 > 1.0$ s). Nevertheless, no notable further response reduction is observed when larger values of hysteretic or viscous damping are used.

**Introduction**

The concept of coupling the lateral response of a moment resisting frame with a rigid core system goes back to the early works of Paulay [1] and Fintel [2]. With this design, interstory drift demands are reduced at the expense of transferring appreciable shear-forces and bending moments at the foundation of the rigid core wall. In the early 1970s a new concept for seismic protection, by modifying the earthquake response of structures with specially designed supplemental devices, was brought forward in the seminal papers by Kelly et al. [3] and Skinner et al. [4] and was implemented in important structures that were under design at that time such as the South Rangitikei Rail Bridge, [5–7] the Union House Building in Auckland [8] and the Wellington Central Police Station in Wellington [9], New Zealand. Despite its remarkable originality and technical merit, the paper by Kelly et al. 1972 did not receive the attention it deserved, and it was some two decades later that the PRESSS Program [10,11] reintroduced the concept of uplifting and rocking of the joint shear wall system [12,13]. Given that damping during impact as the wall alternate pivot points is low, the idea of introducing supplemental energy dissipation devices in structural systems coupled with rocking walls received revived attention [14–18]. In view of the recent findings, this paper examines the contribution of viscous and hysteretic dampers to the response of a yielding frame coupled with a rocking wall shown in Figure 1.

---

[1] Postdoc, Dept. of Civil Eng., Univ. of Southern California, Los Angeles, CA 90089 (email: aghaghol@usc.edu)
[2] Professor, Dept. of Civil Eng., Southern Methodist University, Dallas, TX 75205

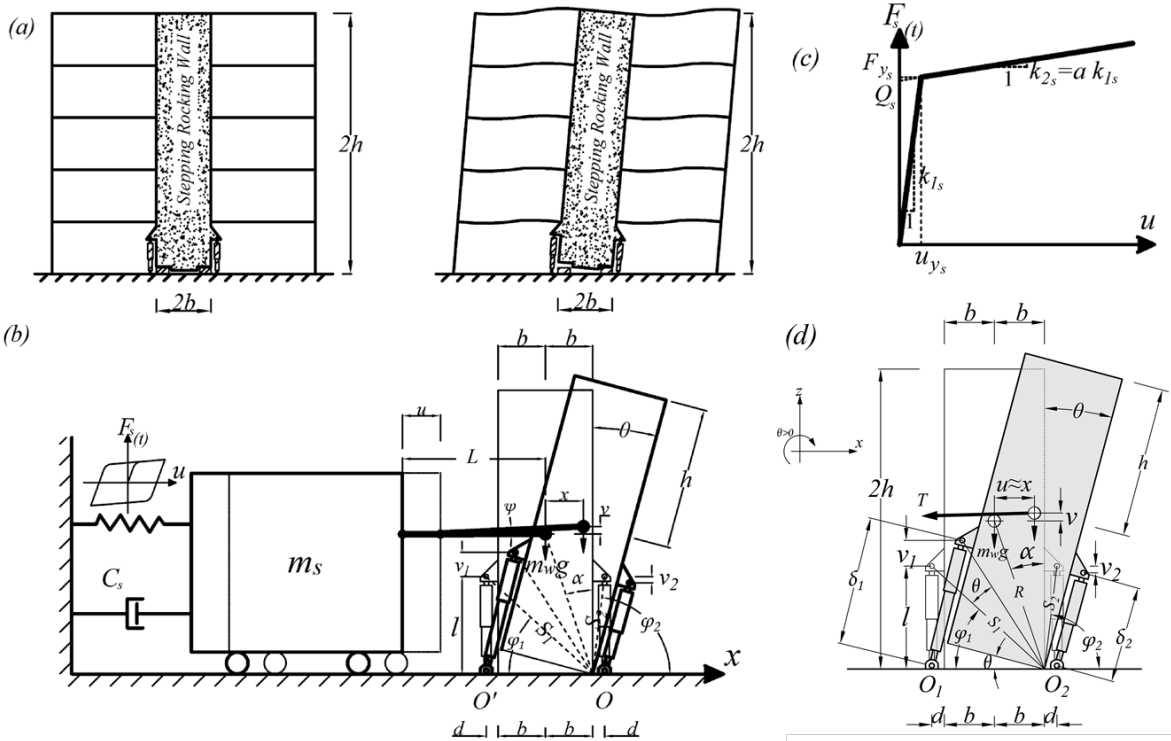

FIGURE 1 (a) Moment-resisting frame with a stepping rocking wall with dampers (b) A SDOF idealization of the yielding frame-rocking-wall system with a yielding oscillator coupled with a rocking wall with supplemental dampers. (c) Bilinear behavior of the yielding SDOF oscillator shown. (d) Geometric quantities pertinent to the dynamic analysis of a rocking wall with additional energy dissipators.

**Dynamics of a Yielding Oscillator Coupled to a Rocking Wall with Supplemental Damping**

With reference to Fig. 1(b), this study examines the dynamic response of a yielding single-degree-of-freedom (SDOF) structure, with mass, $m_s$, pre-yielding stiffness, $k_1$, post-yielding stiffness, $k_2$ and strength, $Q$ that is coupled with a free-standing stepping rocking wall of size, $R = \sqrt{b^2 + h^2}$, slenderness, $\tan \alpha = b/h$, mass mw and moment of inertia about the pivoting (stepping) points $O$ and $O'$, $I = 4/3 m_w R^2$. Vertical energy dissipation devices are mounted to the rocking wall at a distance, $d$, from the pivoting points of the wall as shown in Figures 1(b) and (d). In the interest of simplicity, it is assumed that the arm with length $L$, that couples the motion is articulated at the center mass of the rocking wall at a height, h from its foundation.

With reference to Figure 1(b) dynamic equilibrium of the mass $m_s$ gives:

$$m_s(\ddot{u} + \ddot{u}_g) = -F_s(t) - c_s \dot{u} + T \tag{1}$$

Following a similar procedure as it described in [19,20], the equation of motion for positive and negative rotation ($\theta$) of the stepping rocking wall can be written as:

$$(\tfrac{4}{3} + \sigma \cos^2(\alpha - \theta))\ddot{\theta} + \sigma \cos(\alpha - \theta)\left[a_s \omega_{1s}^2 (\sin\alpha - \sin(\alpha - \theta)) + 2\xi_s \omega_{1s} \dot{\theta}\cos(\alpha - \theta) + \dot{\theta}^2 \sin(\alpha - \theta) + (1 - a_s)\omega_{1s}^2 \frac{u_{ys}}{R} z_s(t)\right] \tag{2}$$

$$= -\frac{g}{R}\left[(\sigma+1)\frac{\ddot{u}_g}{g}\cos(\alpha-\theta) + \sin(\alpha-\theta) + \frac{F_{d_1}}{m_w g}\frac{r_1}{R} + \frac{F_{d_2}}{m_w g}\frac{r_2}{R}\right] \quad for \quad \theta \geq 0$$

$$(\tfrac{4}{3} + \sigma \cos^2(\alpha + \theta))\ddot{\theta} - \sigma \cos(\alpha + \theta)\left[a_s \omega_{1s}^2 (\sin\alpha - \sin(\alpha + \theta)) - 2\xi_s \omega_{1s} \dot{\theta}\cos(\alpha + \theta) + \dot{\theta}^2 \sin(\alpha + \theta) - (1 - a_s)\omega_{1s}^2 \frac{u_{ys}}{R} z_s(t)\right]$$

$$= -\frac{g}{R}\left[(\sigma+1)\frac{\ddot{u}_g}{g}\cos(\alpha+\theta) - \sin(\alpha+\theta) + \frac{F_{d_1}}{m_w g}\frac{r_1}{R} + \frac{F_{d_2}}{m_w g}\frac{r_2}{R}\right] \quad for \quad \theta < 0$$

in which $\sigma = m_s/m_w$, $\xi_s$ = preyielding viscous damping ration of the SDOF oscillator ($\xi_S = 3\%$), $\alpha_S$ is the

post- to pre- yielding stiffness ration in hysteretic Bouc-Wen model [21,22], and $F_d$ is the damping force generated by the additional damping devices attached to the sides of the rocking-wall [19,23,24].

**Earthquake Spectra of a Yielding Oscillator Coupled to a Rocking Wall with Supplemental Damping**

The effect of supplemental damping, either hysteretic or viscous along the sides of a stepping rocking wall coupled to a medium-to-high rise, yielding building is investigated with the generation of inelastic response spectra. Figures 2 and 3 (top and bottom) plots displacement response spectra of the yielding SDOF oscillator coupled to a rocking wall with vertical hysteretic dampers or similarly defined strength viscous dampers (for detailed calculations refer to [19]) appended to the pivot corners of the rocking wall with strength equal to 20% and 50% of the yielding strength of the structure $Q_s = 0.08 m_s g$, postyield-to-preyield stiffness ratio equal to $a_d = 2.5\%$ and the mass ratio $\sigma = m_s/m_w = 10$.

Figure 2 shows that when the input ground motion is the 1994 Newhall record, the vertical hysteretic dampers further suppress the inelastic displacements. Figure 2 (top) also plots the peak angular velocity, $\dot{\theta}_{max}$, of the rocking wall with the scale shown on the right of the plots. Clearly, as the preyielding period, $T_1$, of the frame structure increases, the peak angular velocity decreases. For each value of the preyielding period of the yielding oscillator appearing along the horizontal axis of the spectra, the value, $\dot{\theta}_{max}$, is used to calculate the equivalent viscous damping $c_1 = \epsilon Q_s/(2b\dot{\theta}_{max})$ that is needed to compute the corresponding spectra where the supplemental damping at the pivot corners of the rocking wall are linear viscous dampers. Figure 2 (bottom row) plots displacement response spectra of the yielding SDOF oscillator coupled to a rocking wall with vertical linear viscous dampers appended at the pivot corners of the rocking wall (d=0) with damping constant $c_1 = \epsilon Q_s/(2b\dot{\theta}_{max})$ in which $\dot{\theta}_{max}$ is offered in Figure 2 (top row). Similarly, when the input ground motion is the 1994 Newhall record, the viscous dampers further suppress the inelastic displacements.

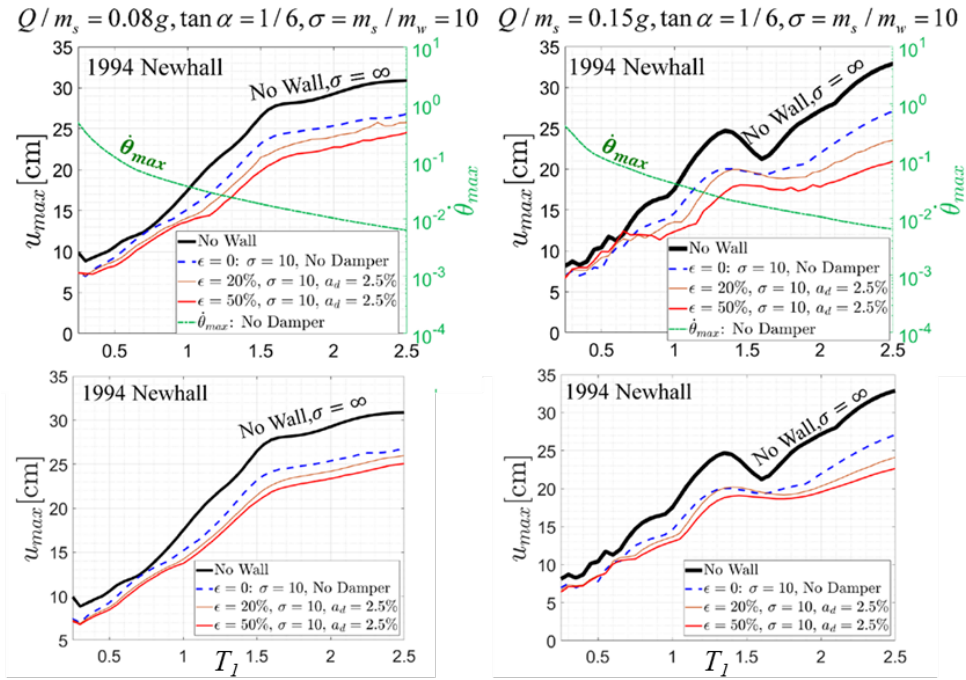

FIGURE 2 Peak response of SDOF yielding oscillator coupled with a stepping wall with zero-length supplemental hysteretic (top row) and viscous dampers (bottom row) appended at the pivoting points ($d = 0$) when excited by the Newhall/360 ground motion recorded during 1994 Northridge, California earthquake. Figures on the left correspond to a SDOF yielding oscillator with strength of $Q_s/m_s = 0.08g$, whereas for the figures on the right, $Q_s/m_s = 0.15g$.

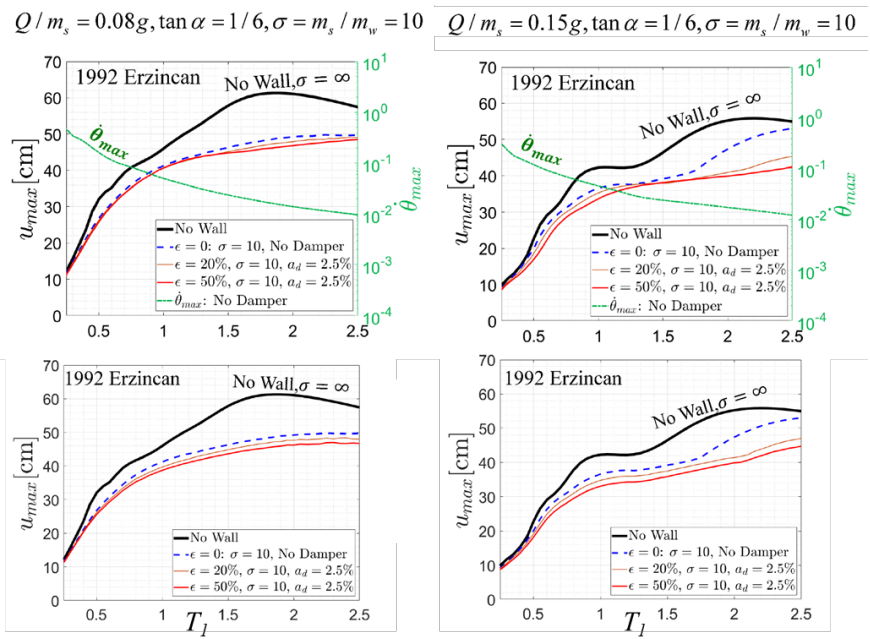

FIGURE 3 Peak response of SDOF yielding oscillator coupled with a stepping wall with zero-length supplemental hysteretic (top row) and viscous dampers (bottom row) appended at the pivoting points ($d = 0$) when excited by the Erzincan NS ground motion recorded during the 1992 Erzincan, Turkey earthquake. Figures on the left correspond to a SDOF yielding oscillator with strength of $Q_s/m_s = 0.08g$, whereas for the figures on the right, $Q_s/m_s = 0.15g$.

There are situations where the structural response exceeds the structural response without dampers being appended at the pivot corners of the rocking wall. This "counter intuitive" finding should not be a surprise since it has been observed to also happen on the rocking response of solitary columns with supplemental damping [19,20] and results from the way that inertia, gravity, and damping forces combine. Similar trends can be observed in Figure 3 when the yielding SDOF system is coupled to the damped rocking wall when subjected to the Erzincan NS ground motion recorded during the 1992 Erzincan, Turkey earthquake.

## Conclusions

The need to ensure uniform interstory drift distribution in medium-to-high rise buildings when subjected to earthquake shaking has prompted a growing interest in coupling the lateral response of moment resisting frames to strong, heavy rocking walls. Given that the coupling with a strong rocking wall, enforces a deformation pattern of the yielding structural system that resembles a first mode, the analysis adopted a single-degree-of-freedom idealization to perform a response analysis.

This paper investigates the dynamic response of a yielding SDOF oscillator coupled to a stepping rocking wall with supplemental damping (either hysteretic or viscous) along its sides. The full nonlinear equations of motion are derived, and the study presents a parametric analysis of the inelastic system in terms of inelastic response spectra and reaches the following conclusions:

The participation of the stepping rocking wall suppresses invariably peak inelastic displacement; as has been shown in previous studies. In contrast, the effect of supplemental damping along the sides of the rocking wall is marginal for structures with preyielding periods lower that $T_1=1.0$ s and occasionally the damped response exceeds the undamped response. Whenever the damped response exceeds the undamped response, the exceedance is marginal.

The paper shows that supplemental damping along the sides of the rocking wall may have some limited beneficial effects for structures with longer preyielding periods (say $T_1>1.0$ s). Nevertheless, no notable further response reduction is observed when larger values of hysteretic or viscous damping are used.